\documentclass[12pt]{article}
\usepackage{color}
\usepackage[dvips]{graphicx}
\usepackage{dcolumn}
\usepackage{amsmath}
\usepackage{url}

\begin{document}

\Large
\begin{center}
	Binding Energies for Successive Addition Reaction of $^\bullet$OH with 
	C$_{60}$: A Laboratory for Testing Frontier Molecular Orbital Theory
\end{center}
\normalsize

\vspace{0.5cm}

\noindent
Abraham PONRA\\
{\em Department of Physics, Faculty of Science, University of Maroua, P.O.\ Box 814, Maroua, CAMEROON\\
	e-mail: abraponra@yahoo.com}

\vspace{0.5cm}

\noindent
Anne Justine ETINDELE\\
{\em Department of Physics, Higher Teachers Training College, University of Yaounde I,
	P.O.\ Box 47, Yaounde, CAMEROON\\
	e-mail: anne.etindele@univ-yaounde1.cm}

\vspace{0.5cm}

\noindent
Ousmanou MOTAPON\\
{\em Department of Physics, Faculty of Science, University of Maroua, P.O.\ Box 814, Maroua, CAMEROON,\\
	Department of Physics, Faculty of Science, University of Douala, P.O.\ Box 24157, Douala, CAMEROON\\
	e-mail: omotapon@univ-douala.com} 

\vspace{0.5cm}

\noindent
Mark E.\ CASIDA\\
{\em Laboratoire de Spectrom\'etrie, Interactions et Chimie th\'eorique 
	(SITh),
	D\'epartement de Chimie Mol\'eculaire (DCM, UMR CNRS/UGA 5250),
	Institut de Chimie Mol\'eculaire de Grenoble (ICMG, FR2607), 
	Universit\'e Grenoble Alpes (UGA)
	301 rue de la Chimie, BP 53, F-38041 Grenoble Cedex 9, FRANCE\\
	e-mail: mark.casida@univ-grenoble-alpes.fr} 

\vspace{0.5cm}

\noindent
{\color{magenta}   
	Date of Publication: \today \hspace{0.1cm} (MS 5.00)
}                  

\vspace{0.5cm}
\begin{abstract}

Buckminsterfullerene C$_{60}$ is proposed as a radical sponge for scavenging
reactive oxygen species such as the hydroxyl radical $^\bullet$OH.  Reaction
energies are calculated using density-functional theory at the B3LYP-D4/def2-SVP
level for successive gas-phase addition reactions of $^\bullet$OH to 
C$_{60}$ up through and including six hydroxyl radicals.  In total, 285 
reactions were investigated yielding minimum energy structures, each of 
which is not necessarily the 
lowest energy conformer but is estimated to be within 5 kcal/mol of the
lowest energy conformer for each new addition.  We confirm that the lowest
energy isomers form a belt of hydroxyl groups around the equator of C$_{60}$,
but ask the question of what governs the relative stability of subtitutions
at different carbons?  Factors concerning regioselectivity are analyzed in 
terms of conceptual density-functional theory, frontier molecular orbital
theory, charge and spin densities, based upon Mulliken population analysis.  
The complexity of applying such an analysis to such large quasi-degenerate
and often open-shell systems results in noisy data but this is adequately
off-set by the quantity of data examined which allow the identification of
clear statistical trends.
We confirm that $^\bullet$OH is an electrophilic radical whose successive 
reaction with C$_{60}$ is under both charge and orbital control.  This is 
seen to be especially the case for addition to odd 
$^\bullet$C$_{60}$(OH)$_{2m+1}$ fullerenols, but is also seen from a
Fukui function and dual descriptor analysis for even C$_{60}$(OH)$_{2m}$
fullerenols. Of particular interest is the ability of the condensed
radical Fukui function $f^0$ to provide information about the reactivity
of even C$_{60}$(OH)$_{2m}$ fullerenols with $^\bullet$OH also when the
spin density is zero, and the observation that the interpretation of the
sign of the dual descriptor changes depending upon whether a spin-restricted
calculation is being performed for even C$_{60}$(OH)$_{2m}$
fullerenols or a spin-unrestricted calculation is being performed for
odd $^\bullet$C$_{60}$(OH)$_{2m+1}$ fullerenols.

\end{abstract}

\section{Introduction}

Since its discovery in 1985 \cite{KHO+85} , the nanocarbon material 
buckminsterfullerene C$_{60}$ has been the subject of extensive research and 
applications. One dream is that suitably functionalized C$_{60}$ could be
rendered both soluble and useful for medical applications \cite{E18}.  A
particularly simple class of water-soluble functionalized C$_{60}$ molecules
consists of the fullerenols C$_{60}$(OH)$_n$ 
\cite{CUS92,LTO+93,CBW+96,CMW+00,RG04,KMT+08,FR09,RTS+11,SMM+11,UYA+14,%
	SCP+16,AKMM17,KT17,VGG18,WGZ18,KSV+19,ZSM+20}.
Because of the large number of possible isomers, this important class of 
compounds provides a ``laboratory'' for exploring theories of chemical 
reactivity.  Previous theoretical work has focused mainly on finding the
lowest energy isomers \cite{CMW+00,RG04,HZJ+11} --- i.e., those with the
highest bond dissociation energies (BDE) --- with much less done to 
\marginpar{\color{blue} BDE}
rationalize regioselectivity for particular binding sites 
\cite{MML10,RTS+11,MFMR14,VGG18,CR19,HN20}.  Traditional explanations of regioselectivity 
include sterochemistry, charge control, and orbital control.  Over time, these
initially qualitative ideas have become increasingly precise and quantifiable
thanks to various advances in theoretical chemical physics.  The result
is an increasingly well-defined set of reactivity indices (RIs) which
\marginpar{\color{blue} RI}
could, for example, be implemented in computer programs to sift through the
large quantities of output data from high performance computing in order to 
seek candidate molecules for specific applications.  However, finding
the best RIs to characterize a given family of molecules and reactions
is complicated both by competing physical effects and by a certain 
redundance among different RIs.  Here we seek the best RIs from
a reasonable list of different RIs for predicting the lowest energy 
isomers ($^\bullet$)C$_{60}$(OH)$_n$ resulting from successive gas-phase
reactions of buckminsterfullerene (C$_{60}$) with the hydroxyl radical 
($^\bullet$OH).  [Note that ($^\bullet$)C$_{60}$(OH)$_n$ has an 
unpaired electron only for $n$ odd.]

Buckminsterfullerene has been termed a radical sponge because of its 
ability to react with a large number of radicals \cite{MML92}.  
It has even been commercialized under this name for use in 
cosmetics \cite{Sponge}.  A typical radical reaction is with
the hydroxyl radical $^\bullet$OH to form the fullerenols 
($^\bullet$)C$_{60}$(OH)$_n$.  Fullerenols are attractive because, in
contrast with the parent C$_{60}$, fullerenols are water soluble which
is important for biological applications as radical scavangers.
Syntheses of C$_{60}$(OH)$_n$ fullerenols were already reported
during the first decade after the discovery of C$_{60}$ 
\cite{CUS92,LTO+93,CBW+96}.  These were done either by indirect
processes leading to as many as 20 hydroxyl units attached to the
central fullerene \cite{CUS92,CBW+96} or by a direct process consisting
of the reaction of C$_{60}$ with NaOH in the presence of tetrabutylammoniun 
hydroxide (TBAH) \cite{LTO+93}.  More recent syntheses have resulted
in fullerenols with up to 44 hydroxyl units \cite{KMT+08, SMM+11}.

Of course, this immediately means that we have a horrendous isomer
problem.  There are 60 carbons to which a hydroxide radical may bind.  
Knowing that one of the hydroxides must be placed at carbon 1, this 
gives a maximum of
\begin{equation}
\left( \begin{array}{c} 59 \\ n-1 \end{array} \right) = \frac{59!}{(60-n)! (n-1)!}
\label{eq:intro.1}
\end{equation}
isomers for C$_{60}$(OH)$_n$.  However additional symmetries mean that there
will be fewer isomers in practice.  {\bf Table~\ref{tab:count}} shows how 
the maximum value of Eq.~(\ref{eq:intro.1}) grows as a function of $n$.

\begin{table}
	\caption{Maximum number of isomers of C$_{60}$(OH)$_n$ calculated
		from Eq.~(\ref{eq:intro.1}).
		\label{tab:count}
	}
	\begin{center}
		\begin{tabular}{cc}
			\hline \hline
			$n$ & isomers \\
			\hline
			1 & 1 \\
			2 & 59 \\
			3 & 1534 \\
			4 & 32509 \\
			\hline \hline
		\end{tabular}
	\end{center}
\end{table}

\begin{figure}
	\begin{center}
		\begin{tabular}{cc}
		
			\includegraphics[width=0.5\textwidth]{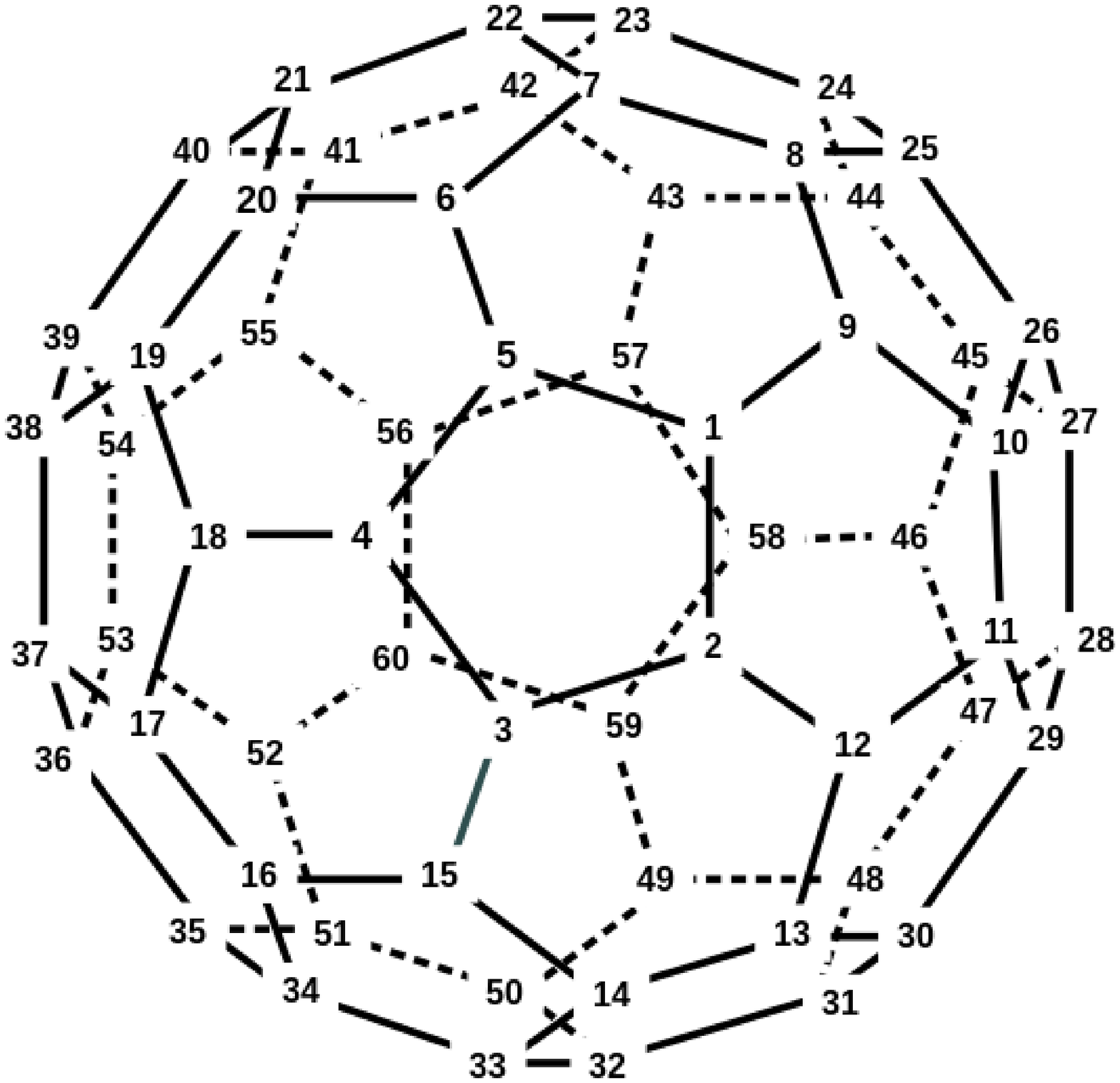}
			&
			\includegraphics[width=0.5\textwidth]{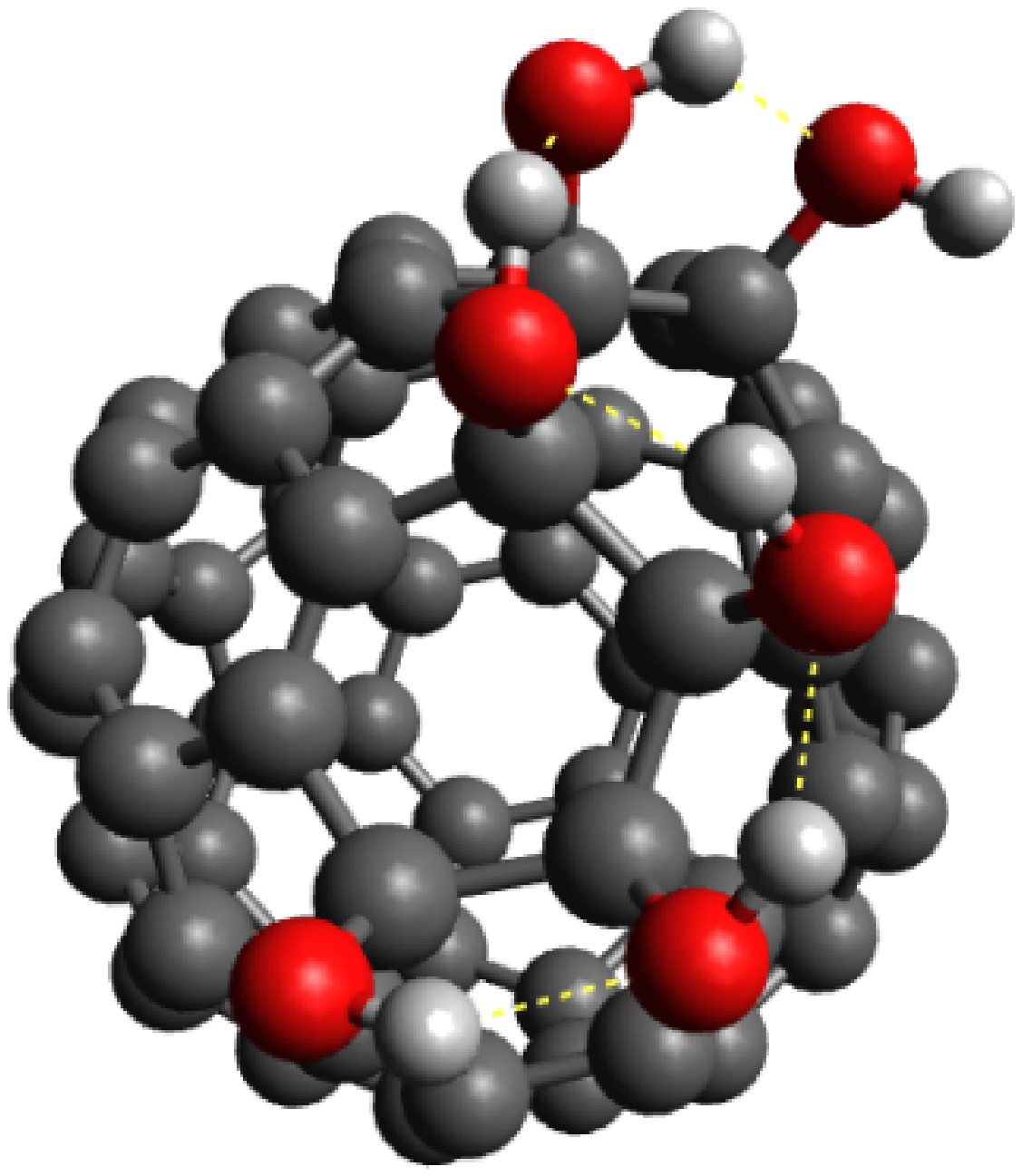}
		\end{tabular}
	\end{center}
	\caption{
		Left-hand side: IUPAC numbering \cite{PCM+02}.  Right-hand side:
		1,5,6,9,10,11-C$_{60}$(OH)$_6$, the most stable isomer found in this
		study.
		\label{fig:C60OH6geom}
	}
\end{figure}

Previous theoretical studies using semi-empirical, density-functional theory
(DFT), and {\em ab initio} methods 
\marginpar{\color{blue} DFT}
have focused on determining the most stable isomers \cite{CMW+00,RG04,HZJ+11}.
We know from those studies that the hydroxyls are initially added in such 
a way as to form a belt around the equator of the roughly spherical C$_{60}$.
The results presented here ({\bf Fig.~\ref{fig:C60OH6geom}}) are consistent 
with that prediction.  Note that we follow the IUPAC convention \cite{PCM+02}
throughout when numbering the hydroxyl positions, with the first addition 
occurring at C1.

However, our main concern is with {\em why} the hydroxyl radicals add where 
they add.  Traditional chemical reasoning uses fuzzy qualitative concepts 
such as steric hinderance, inductive and resonance effects, and charge 
and orbital control.  Recent work, notably in conceptual density-functional 
theory (CDFT)
\marginpar{\color{blue} CDFT}
\cite{C99,GDL03,GCC+20,GR21,L22} , has helped to replace these
concepts with well-defined quantitative RIs.  One set of these are the Fukui
functions (Sec.~\ref{sec:FMO}) which are associated with regioselectivity. 
Several studies have calculated Fukui functions for C$_{60}$ and related 
species \cite{MML10,MFMR14,CR19,HN20}.  Particularly notable from the point
of view of the present study is that of Rodr\'{\i}guez-Zavala {\em et al.}\ 
\cite{RTS+11} who studied the sequential attachment of OH groups to C$_{82}$
using two RIs, namely condensed Fukui functions and visualization
of frontier molecular orbitals (FMOs) \cite{RTS+11}. But Fukui functions
\marginpar{\color{blue} FMO}
are not the only relevant descriptors.  Haghgoo and Nekoei \cite{HN20} have 
noted that high Fukui function values do not necessarily indicate the most 
active sites for electrophilic and nucleophilic attack and they propose to 
add to this the calculation of natural bond orbital (NBO) charges \cite{HN20}. 
In the context of FMO theory (Sec.~\ref{sec:FMO}), charges, and Fukui 
functions correspond, respectively, to charge and orbital control.
A third RI to consider for species with an odd number of electrons is
the spin density, $\rho_\uparrow - \rho_\downarrow$.

The present work examines the relative predictive value of charges, 
Fukui functions, and spin-densities for finding the stability of 
different isomers of ($^\bullet$)C$_{60}$(OH)$_n$ ($n$ = 0-5) obtained 
by successive addition of $^\bullet$OH to C$_{60}$.  Of particular
interest is how electrophilic $^\bullet$OH behaves when it comes to
regioselectivity on the radical sponge ($^\bullet$)C$_{60}$(OH)$_n$?
Radicals are classified as nucleophilic or electrophilic \cite{DSV+07,binkley} ,
and the hydroxyl radical is normally classified as strongly electrophilic
\cite{MIN+02} , which makes it an electron acceptor.  But the radical
sponge C$_{60}$ is also an electron acceptor.  We will use the dual
descriptor to decide the extent to which
($^\bullet$)C$_{60}$(OH)$_n$ acts as an electron donor or an electron
acceptor with respect to regioselectivity of reaction with $^\bullet$OH.
Our conclusions, which depend upon $n$, are backed up by calculation and
analysis of 286 BDEs.

This paper is organized as follows: FMO theory and CDFT are reviewed in 
the next section. Section 3 provides computational details.
This is followed by an analysis of our results in Section 4
and a concluding discussion in Section 5.  Additional 
information is provided as Supplementary Material.

\section{Conceptual Density Functional Theory}

This section provides enough review of the CDFT literature to keep this
paper relatively self-contained.  It also addresses issues specific to
the application in this article. 

The basic idea of CDFT is an expansion of the total energy in derivatives with
respect to the number of electrons and the external potential thereby providing
local information about the how energy changes when atoms approach each other and
electrons are transferred \cite{APP06,GCC+20,GR21,PY84,PY89,GDL03,YCDG12,SZS+14,L22}.
It has even been suggested that this CDFT expansion may be used to guide potential
energy searches \cite{GR21}. CDFT quantities are often evaluated in a finite
\marginpar{\color{blue} FDA}
difference approximation (FDA).  FMO theory arises as a further frozen orbital 
approxiation (FOA).  This is important because it shows how CDFT provides a more
\marginpar{\color{blue} FOA}
rigorous formulation of FMO theory but also because FMO theory provides an 
intepretation for CDFT quantities.  It is therefore wise to begin with a brief 
review of FMO theory.  This review is inseparable from Pearson's hard 
and soft Lewis acids and bases (HSAB) theory of chemical reactivity 
\marginpar{\color{blue} HSAB}
\cite{P63,P68a,P68b,PP83,P05} since an important use of both modern FMO theory and 
CDFT has been in justifying HSAB theory.

The fact that Fukui \cite{FYS52} and Hoffmann were jointly awarded the 1981 Nobel prize in chemistry 
for their work on FMO theory is an indication of just how important and ingrained FMO theory 
is in chemistry.  Textbooks have been written on the subject \cite{F76,A07} and it is
difficult to publish criticism \cite{D89}.  Nevertheless some review of the
subject is in order, if only to emphasize some of the finer points that are
needed for the present paper. Klopman and Salem presented a particularly 
well-known explanation of FMO theory and of the HSAB principle.
Klopman \cite{K68} and Salem \cite{S68a,S68b} considered the reaction of two
molecules A and B.  They used semiempirical theory to decompose the bond energy
into three terms,
\begin{equation}
E(\mbox{A-B}) - [E(\mbox{A})+E(\mbox{B})] = \Delta E = E_{RCF} + E_{Coul} + E_{FMO} \, .
\label{eq:FMO.1}
\end{equation}
Here $E_{RCF}$ represents the first-order repulsion between doubly-filled orbitals in A
and doubly-filled orbitals in B.  It represents steric interactions and is generally 
larger in magnitude than the other two terms.  However this first term is often
relatively constant during many chemical reactions, so that reaction paths are 
then mainly determined by the other two terms.  The first of these other two terms,
$E_{Coul}$ represents the electrostatic Coulomb repulsion energy.  When it dominates,
as would be expected to be the case for example for the reaction between an anion and
a cation, then we say that the reaction is under charge control.  Equivalently, domination
of this term corresponds to the reaction of a hard Lewis acid with a hard Lewis base.
It was often thought that the last term $E_{FMO}$ dominates in reactions between two 
neutral molecules, in which case the reaction is said to be orbital controled.  This
is also the case of the reaction with of a soft Lewis acid with a soft Lewis base.
The dominant part of $E_{FMO}$ is often taken to be,
\marginpar{\color{blue} HOMO (H)}
\begin{equation}
E_{FMO} \approx 
\frac{\vert \langle \psi_{H,A} \vert \hat{F} \vert \psi_{L,B} \rangle \vert^2}
{\epsilon_{H,A} - \epsilon_{L,B}} \, ,
\label{eq:FMO.2}
\end{equation}
where $\hat{F}$ is the orbital hamiltonian, $H,A$ refers to the highest occupied
molecular orbital (HOMO or H) of A, 
and $L,B$ refers to the lowest unoccupied molecular orbital (LUMO or L) of B.   
\marginpar{\color{blue} LUMO (L)}
Then $E_{FMO}$ will be largest when the overlap between $\psi_{H,A}$ and $\psi_{L,B}$ 
is maximized and the two energies $\epsilon_{H,A}$ and $\epsilon_{L,B}$ are close.

Let us now turn to CDFT.  By analogy with thermodynamics, the chemical potential 
$\mu$ is defined by
\begin{equation}
\mu = \left( \frac{\partial E}{\partial N} \right)_v  \, .
\label{eq:FMO.3}
\end{equation}
Pearson's hardness is given a more rigorous definition as,
\begin{equation}
\eta = \left( \frac{\partial \mu}{\partial N} \right)_v 
= \left( \frac{\partial^2 E}{\partial N^2} \right)_v \, .
\label{eq:FMO.4}
\end{equation}
Notice how this is really a second derivative.  The chemical potential and hardness 
are global quantities.  Nonlocal quantities arise by taking derivatives with respect
to the external potential.  The simplest of these is the density itself,
\begin{equation}
\rho(\vec r) = \left( \frac{\delta E}{\delta v(\vec r)} \right)_N \, .
\label{eq:FMO.5}
\end{equation}
The Fukui function \cite{PY84} ,
\begin{equation}
f(\vec{r}) = \left( \frac{\partial \rho(\vec r)}{\partial N} \right)_v \, ,
\label{eq:FMO.6}
\end{equation}
and the dual descriptor (DD) \cite{MGT05,MGT06,M15} ,
\marginpar{\color{blue} DD}
\begin{equation}
\Delta f(\vec{r}) = \left( \frac{\partial f(\vec r)}{\partial N} \right)_v
= \left( \frac{\partial^2 \rho(\vec r)}{\partial N^2} \right)_v \, .
\label{eq:FMO.7}
\end{equation}

These simple definitions are complicated by the presence of a particle number discontinuity
for integer $N$ \cite{PL83,SS83}.  So distinctions must be made between left derivatives
for $N-\delta$ designated by a superscript $^-$, right derivatives for $N+\delta$ designated
by a superscript $^+$, and the average of the left and right derivatives designated by
a superscript $^0$.  For the chemical potential,
\begin{eqnarray}
\mu^- & = & -I \nonumber \\
\mu^+ & = & -A \nonumber \\
\mu^0 & = & -(I+A)/2 = -\chi \, ,
\label{eq:FMO.8}
\end{eqnarray}
where $I$ is the ionization potential, $A$ is the electron affinity, and $\chi$ is the
Mulliken electronegativity.  Similarly we can define the corresponding Fukui function quantities.
The FDA allows us to obtain some insight into the CDFT quantities already mentioned:
\begin{eqnarray}
\eta & \approx & \mu^+-\mu^- = I - A \nonumber \\
f^-(\vec r) & \approx & \rho_N(\vec r) - \rho_{N-1}(\vec r) \nonumber \\
f^+(\vec r) & \approx & \rho_{N+1}(\vec r) - \rho_N(\vec r) \nonumber \\
f^0(\vec r) & \approx & \frac{\rho_{N+1}(\vec r) - \rho_{N-1}(\vec r)}{2} \nonumber \\
\Delta f(\vec r) & \approx & f^+(\vec r) - f^-(\vec r) 
\approx \rho_{N+1}(\vec r) - 2\rho_N(\vec r) + \rho_{N-1}(\vec r) \, .
\label{eq:FMO.9}
\end{eqnarray}

However the connection to FMO theory is only made by making the FOA in addition to the FDA.
The result depends upon whether our molecule is closed-shell with all electrons paired 
or has one singly occupied
\marginpar{\color{blue} SOMO (S)}
molecular orbital (SOMO or S).  This, of course, is important for us when we study the
reaction,
\begin{equation}
\mbox{C$_{60}$(OH)$_n$} + \mbox{OH} \rightarrow \mbox{C$_{60}$(OH)$_{n+1}$} \, 
\label{eq:FMO.10}
\end{equation}
because $^\bullet$OH is a neutral radical.  When $n=2m$ is even ($m=0$) is allowed,
\begin{equation}
\mbox{C$_{60}$(OH)$_{2m}$} + \mbox{$^\bullet$OH} \rightarrow 
\mbox{$^\bullet$C$_{60}$(OH)$_{2m+1}$} \, ,
\label{eq:FMO.11}
\end{equation}
then the FOA gives
\begin{eqnarray}
f^-(\vec{r}) & \approx & \vert \psi_H(\vec{r}) \vert^2 \nonumber \\ 
f^+(\vec{r}) & \approx & \vert \psi_L(\vec{r}) \vert^2 \nonumber \\
f^0(\vec{r}) & \approx & 
\left( \vert \psi_L(\vec{r}) \vert^2 + \vert \psi_H(\vec{r}) \vert^2 \right)/2 
\nonumber \\
\Delta f(\vec{r}) & \approx & 
\vert \psi_L(\vec{r}) \vert^2 - \vert \psi_H(\vec{r}) \vert^2  \, .
\label{eq:FMO.12}
\end{eqnarray}
We now see that the Fukui functions are telling us about regioselectivity. 
The parts of the molecule where the HOMO density is large will be regioselective
for attack by electron acceptors (electrophiles) and this is the same for the 
``nucleophilic'' Fukui function $f^-$.  Similarly the parts of the molecule where 
the LUMO density is large will be regioselective for electron donors (such as 
nucleophiles) and this is the same for the ``electrophilic'' Fukui function $f^+$.
As a radical may either donate or accept electrons, then the ``radicalphilic''
Fukui function $f^0$ is said to be indicative of regioselectivity for radicals.  
The dual descriptor $\Delta f$ will be positive for electron-accepting regions 
and negative for electron-donating regions.

In contrast, when $n=2m+1$ is odd ($m=0$ is allowed), then,
\begin{eqnarray}
f^-(\vec{r}) & \approx & \vert \psi_S(\vec{r}) \vert^2 \nonumber \\
f^+(\vec{r}) & \approx & \vert \psi_S(\vec{r}) \vert^2 \nonumber \\ 
f^0(\vec{r}) & \approx & \vert \psi_S(\vec{r}) \vert^2 \nonumber \\
\Delta f(\vec{r}) & \approx & 0 \, ,
\label{eq:FMO.13}
\end{eqnarray}
This makes it look as though radicals are equally electron accepting and
electron donating, which is not entirely satisfactory as radicals are also
classified according to whether they are electrophilic or nucleophilic
\cite{DSV+07,binkley}.  In particular, we would like to know whether 
the regioselectivity of $^\bullet$OH for a given ($^\bullet$)C$_{60}$(OH)$_n$
is more nucleophilic or electrophilic?  This information may also be
obtained from the dual descriptor, but it will be seen later that
the rule for the sign is {\em not} the same as for the $n=2m$ case.

We actually report condensed Fukui functions (CFFs) in this paper.  These are calculated
\marginpar{\color{blue} CFF}
as
\begin{eqnarray}
f^-_I & \approx & q_I(N-1) - q_I(N) \nonumber \\
f^+_I & \approx & q_I(N)-q_I(N+1) \nonumber \\
f^0_I & \approx & \left( q_I(N-1) - q_I(N+1)  \right)/2  \nonumber \\
\Delta f_I & \approx & q_I(N-1) -2q_I(N) + q_I(N+1)   \nonumber \\
\, ,
\label{eq:FMO.15}
\end{eqnarray}
where $q_I(M) = Z_I-\rho_I$ is the charge on atom $I$ of the $M$-electron species 
(rather than the number of electrons, $\rho_I$).  Of course, there are different ways
to assign charges to atoms in a molecule.  We will just use the usual Mulliken charges,
which seems reasonable for present purposes.  

It also seems important for reasons of completeness to add a few caveats as we end this section.  The first caveat
is that our calculations are for systems with near orbital degeneracies.  This is 
exactly the case where it may not always be possible to explain things using only 
the HOMO, LUMO, or SOMO FMOs.  Nevertheless this is what we have done and our results
seem reasonable.  We have also not tried to include any measure of steric hindrance
in our RIs both because we do not think that regioselectivity is going to be determined
by steric hindrance in our problem and because it is difficult to design RIs to
describe steric hindrance (see, however, Refs.~\cite{L07,LRL17,LLY+18}).

\section{Computational Details}

Calculations were carried out using the {\sc Orca} \cite{N12,N18}
quantum chemistry program.  All calculations are B3LYP-D4 density-functional
theory gas phase calculations.  Here D4 represents Grimme's D4 dispersion
correction \cite{CBG17,CEH+19}. The original B3LYP functional \cite{SDCF94} (referred
to as B3LYP/G in {\sc Orca}) used the same three hybrid parameters as Becke's B3PW
functional \cite{B93} but with different exchange and correlation functionals.
{\sc Orca}'s B3LYP functional differs from the B3LYP/G functional in that the
Vosko-Wilk-Nusair parameterization of Ceperley and Alder's homogeneous
electron gas results \cite{VWN80} is used for the local density approximation 
rather than the Vosko-Wilk-Nusair parameterization of the random-phase approximation result
for the homogeneous electron gas \cite{VWN80}.  The primary difference between
the B3LYP and B3LYP/G functionals is a constant shift in the total energy.
We have chosen to use the def2-SVP orbital basis set \cite{WA05}.  A comparison
with B3LYP-D4 calculations carried out with other basis sets is given in the
Supplementary Information (see, Sec.~\ref{sec:suppl}).
\marginpar{\color{blue} B3LYP-D4/def2-SVP}

Each ($^\bullet$)C$_{60}$(OH)$_n$ geometry was optimized by bringing in an $^\bullet$OH
radical from a large distance towards each carbon.  No significant activation energies
were noticed for these gas phase reactions.  The final geometries were optimized 
in each case and a vibrational analysis was carried out to make sure that the geometries 
are true minima.  However no attempt was made to investigate all possible conformers.  
In order to obtain an idea of the size of possible errors due to neglect of other conformers,
rotation of OH groups was carried out in C$_{60}$(OH)$_2$  to estimate how much energies 
might change due to the making and breaking of hydrogen bonds.  
We found energy variations of as much as about 5 kcal/mol which, though hardly
negligible compared to the BDEs studied here, is still small enough to
allow us to study the effects which interest us.  In any case, a search
for the lowest energy conformer in all the 285 reactions studied here
would be prohibative.

All charge analysis and calculation of condensed Fukui functions is based upon Mulliken
population analysis.  {\sc ChemCraft} \cite{chemcraft} , 
{\sc Avogadro} \cite{avogadro,HCL+12} , and {\sc MOLDEN} \cite{molden,SNM00} 
were used for visualization purposes.

\section{Results}
We wish to test out some common ideas about chemical reactivity from FMO
and related theories for the successive gas-phase addition reactions,
\begin{eqnarray}
\mbox{C$_{60}$(OH)$_{2m}$} + \mbox{$^\bullet$OH} & \rightarrow &
\mbox{$^\bullet$C$_{60}$(OH)$_{2m+1}$} \nonumber \\
\mbox{$^\bullet$C$_{60}$(OH)$_{2m+1}$} + \mbox{$^\bullet$OH} & \rightarrow &
\mbox{C$_{60}$(OH)$_{2m+2}$} \, .
\label{eq:results.1}
\end{eqnarray}
In each case, $^\bullet$OH is being added to the most stable fullerenol
isomer of the previous addition reaction.  According to the Klopman-Salem 
explanation of FMO theory (reviewed in the previous section), reactivity is 
determined by three terms which represent, respectivity, steric, charge, and 
orbital control.  Steric control seems less important for the 
present reactions, so we focus on charge and orbital control.

\begin{table}
	\caption{FMO energies.
		\label{tab:FMOenergies}
	}
	\begin{center}
		\begin{tabular}{ccccccc}
			\hline \hline
			Molecule & HOMO energy & SOMO energy & LUMO energy \\
			\hline
			$^\bullet$OH                          &  & -8.8 eV$^a$  &  \\
			C$_{60}$                              & -6.2 eV$^b$ &       & -3.48 eV$^c$ \\
			$^\bullet$C$_{60}$OH                  &  & -5.37 eV$^d$ &  \\
			1,9-C$_{60}$(OH)$_2$                  & -5.83 eV$^e$ &          & -3.31 eV$^e$ \\
			1,5,9-$^\bullet$C$_{60}$(OH)$_3$      &  & -5.32 eV$^f$ &  \\
			1,5,6,9-C$_{60}$(OH)$_4$              & -5.70 eV$^g$ &          & -3.30 eV$^g$ \\
			1,5,6,9,10-$^\bullet$C$_{60}$(OH)$_5$ &  & -5.57 eV$^h$ &  \\
			1,5,6,7,9,10,11-C$_{60}$(OH)$_6$      & -5.85 eV$^i$ &          & -3.29 eV$^i$ \\
			\hline \hline
		\end{tabular}
	\end{center}
	$^a$ The NIST webbook lists 13.02 eV as the ionization potential of 
	$^\bullet$OH and 1.83 eV as the electron affinity of $^\bullet$OH.  Taking
	minus the average gives -7.42 eV for the SOMO level.
	\\$^b$ The experimental gas-phase ionization potential gives 
	-7.57 eV \cite{MPCG06} while $GW$ calculations gives -7.76 eV \cite{DCT+15}.
	\\$^c$ The experimental gas-phase {\em adiabatic} electron affinity from 
	low-temperature photoelectron spectroscopy gives -2.68 eV \cite{LWW+13}
	while $GW$ calculations give -2.14 eV \cite{DCT+15}.
	\\$^d$ Minus the average of the calculated gas-phase ionization potential (6.99 eV)
	and the calculated gas-phase electron affinity (3.23 eV) is -5.11 eV.
	\\$^e$ Minus the calculated gas-phase ionization potential is -7.48 eV and
	minus the calculated gas-phase electron affinity is -2.58 eV.
	\\$^f$ Minus the average of the calculated gas-phase ionization potential (6.88 eV)
	and the calculated gas-phase electron affinity (3.05 eV) is -4.97 eV.
	\\$^g$ Minus the calculated gas-phase ionization potential is -7.38 eV and
	minus the calculated gas-phase electron affinity is -2.55 eV.
	\\$^h$   Minus the average of the calculated gas-phase ionization potential (6.94 eV)
	and the calculated gas-phase electron affinity (3.24 eV) is -5.09 eV.
	\\$^i$ Minus the calculated gas-phase ionization potential is -6.94 eV and
	minus the calculated gas-phase electron affinity is -3.24 eV.
\end{table}

\begin{figure}
	\begin{center}
		\includegraphics[width=1.2\textwidth]{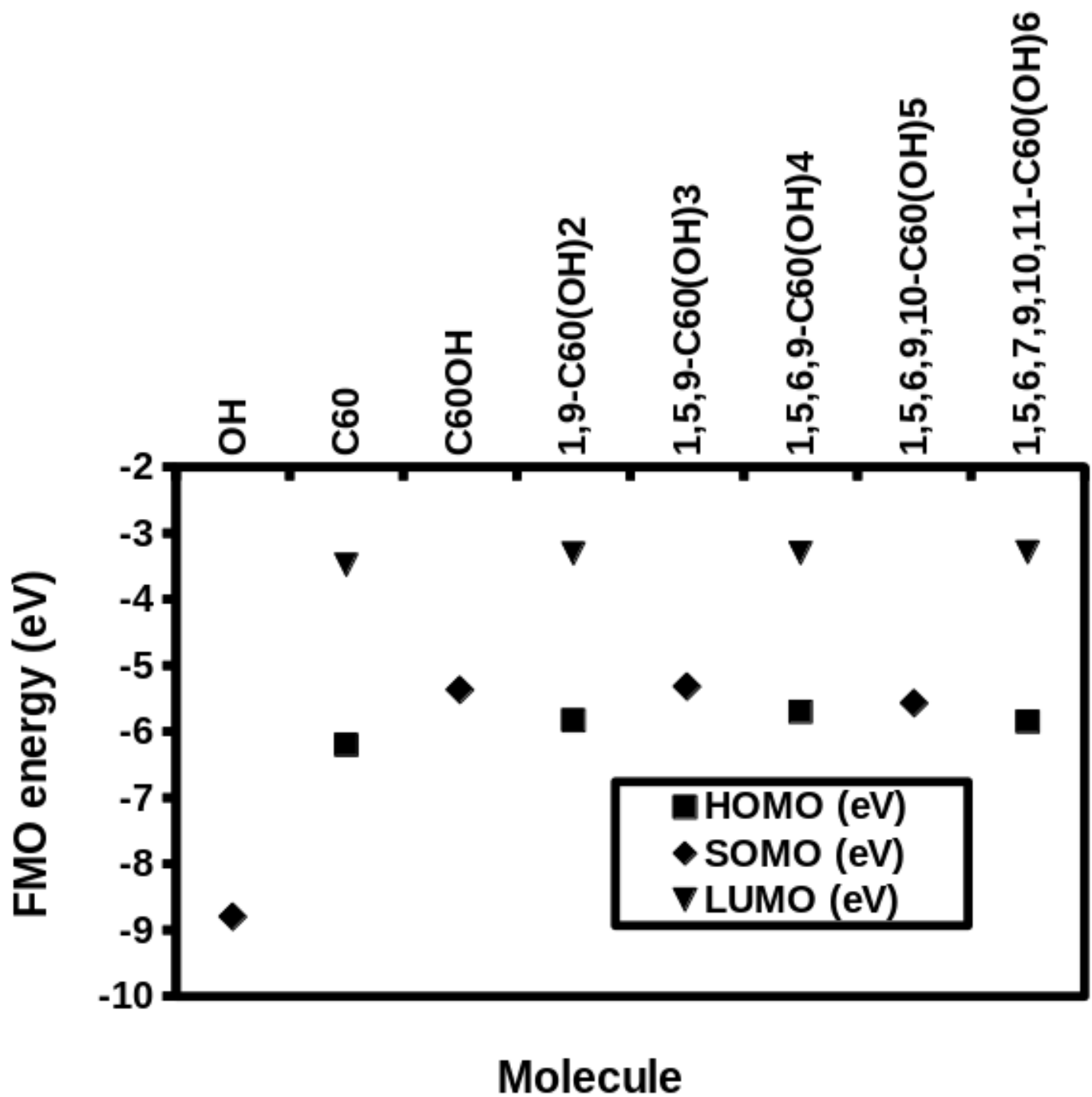}
	\end{center}
	\caption{
		FMO energies for key molecules.  Data from Table~\ref{tab:FMOenergies}.
		\label{fig:FMOenergies}
	}
\end{figure}
\begin{figure}
	\begin{center}
		\includegraphics[width=\textwidth]{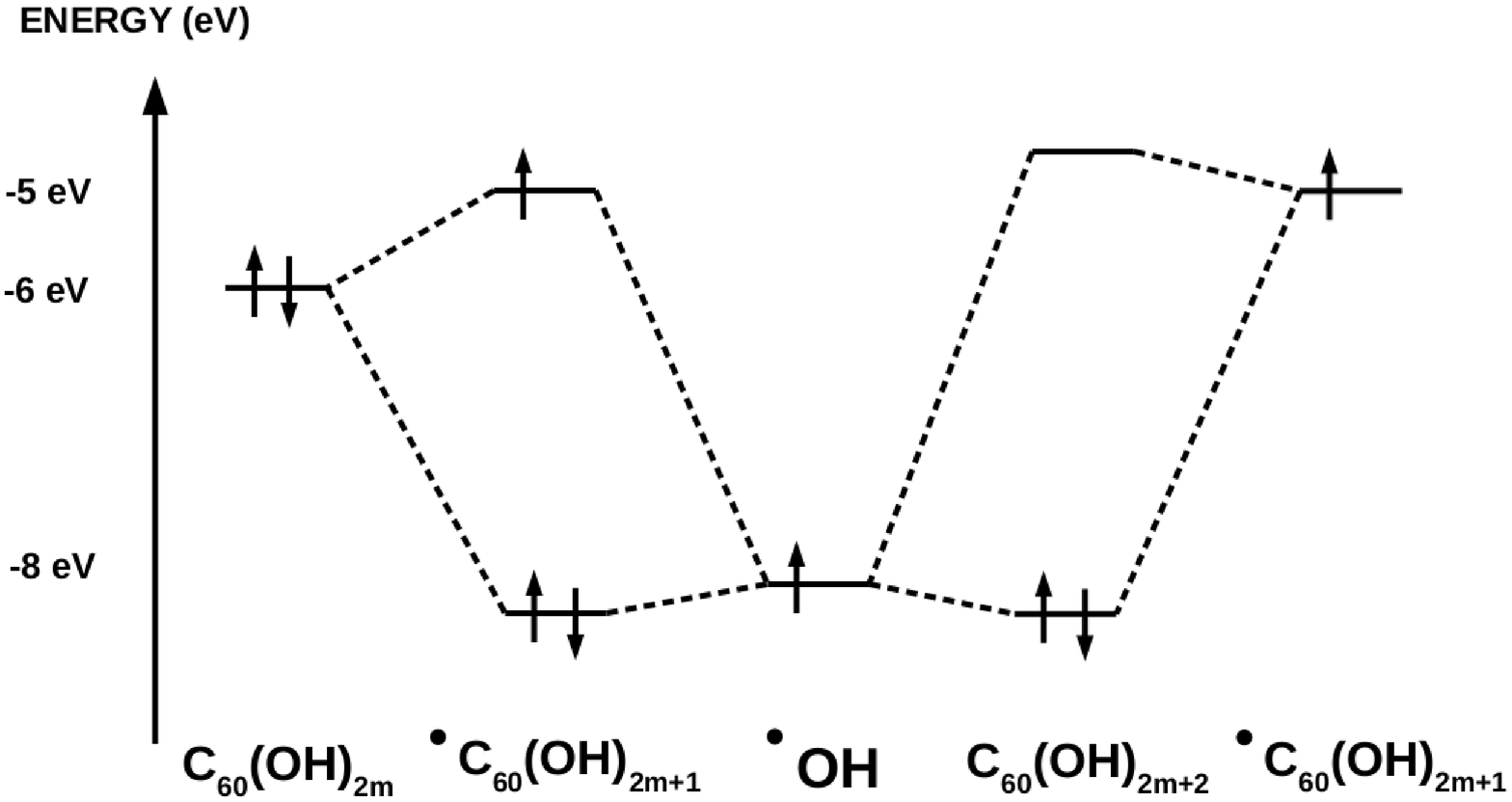}
	\end{center}
	\caption{
		Rough FMO correlation energy diagram.
		\label{fig:FMOcorrelationdiag}
	}
\end{figure}

To discuss charge and orbital control in terms of Klopman-Salem theory, we need
to look at the FMO diagram.  {\bf Figure~\ref{fig:FMOenergies}} shows calculated
FMO energies for the most stable species obtained by successive addition of $^\bullet$OH.
This allows us to construct the rough FMO correlation diagram shown in 
{\bf Fig.~\ref{fig:FMOcorrelationdiag}}.
Note that species with an odd number of electrons are treated by
doing a spin-unrestricted calculation which leads to different energies for spin-up
electrons.  For reasons of conceptual simplicity, we follow a common practice and
obtain the SOMO energy by averaging the energies of the corresponding spin-up and down orbitals.  

\begin{figure}
	\begin{center}
		\includegraphics[width=0.8\textwidth]{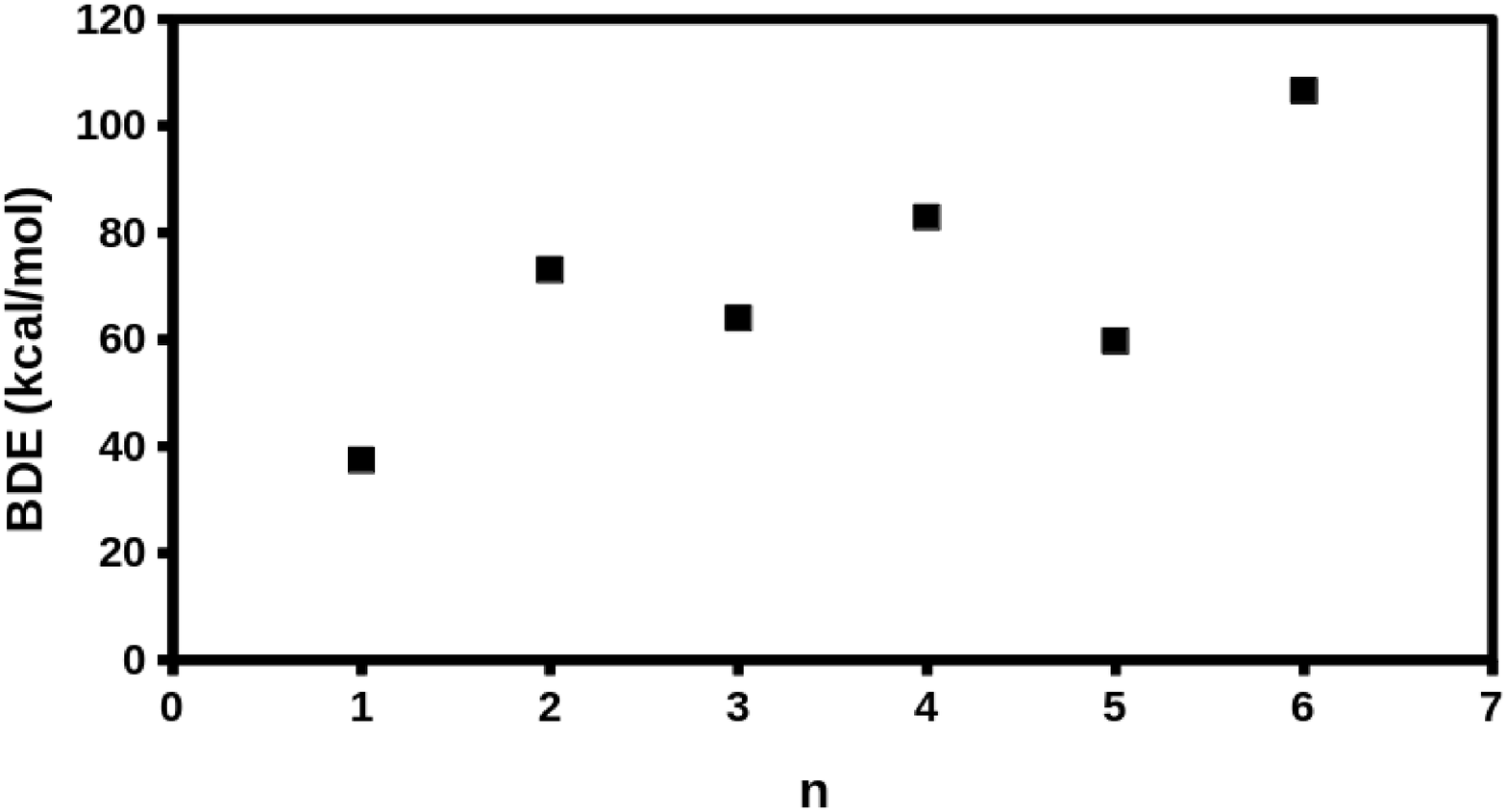}
	\end{center}
	\caption{
		BDE of ($^\bullet$)C$_{60}$(OH)$_n$ as a function of $n$.
		\label{fig:BDEvsN}
	}
\end{figure}
According to the Klopman-Salem theory, the orbital-control term may be neglected
when the denominator in Eq.~(\ref{eq:FMO.2}) is large enough.  When this happens, 
electrons will simply transfer from the HOMO of C$_{60}$(OH)$_{2m}$ to the SOMO 
of $^\bullet$OH to create $^\bullet$C$_{60}$(OH)$_{2m+1}$ or from the SOMO of 
$^\bullet$C$_{60}$(OH)$_{2m+1}$ to the SOMO of $^\bullet$OH to create 
C$_{60}$(OH)$_{2m+2}$.  Making the frequent FMO approximation that sums of orbital
energies may be treated as total energies leads to a predicted bond energy of
about 2 eV or 46 kcal/mol which is about right ({\bf Fig.~\ref{fig:BDEvsN}}).

Let us now turn to the question of regioselectivity.  It is wise to begin
by asking what we might expect.  On the one hand, $^\bullet$OH is expected
to be highly reactive and so perhaps not very regioselective.  On the other
hand, organic chemistry is based upon the idea that functional groups modify 
reactivity by acting locally.  Hence we may expect preferential $^\bullet$OH 
addition close to previously-added hydroxyl groups.  

We might also try to use the concept of a Clar sextet which says that the
most stable structures are expected to be those for which local benzene-like
resonance is possible \cite{S13b}.  However the problem here is one of 
counting Clar sextets, which is a challenging mathematical problem in
and of itself.  For C$_{60}$ alone, Shiu, Lam, and Zhang have predicted
that there are 295 Clar structures \cite{SLZ03}.  This favors the known
icosohedral symmetry structure of C$_{60}$ \cite{ZYL10}.  It also leads
us to suggest that the most stable structures should be the ones with 
the highest number of unsubstituted six-membered carbon rings, which 
could explain the observed preference for equatorial addition 
\cite{CMW+00,RG04,HZJ+11}.

\begin{figure}
	\begin{center}
		\includegraphics[width=1.2\textwidth]{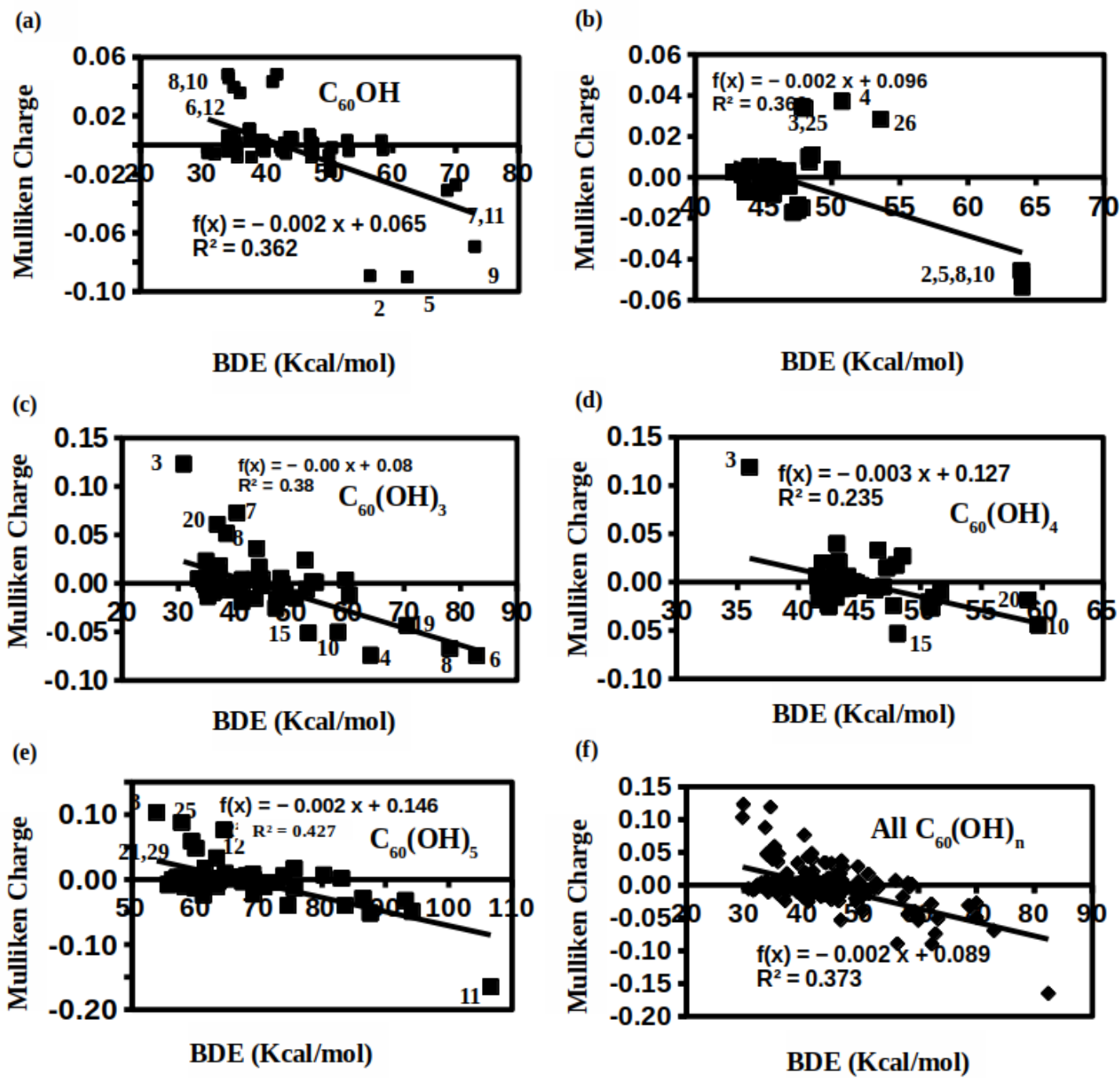}
	\end{center}
	\caption{
		Correlation between carbon Mulliken charges and associated BDEs.
		Some points have been labeled by their carbon number to facilitate discussion.
		The BDE in part (f) is shifted using the statistical data in 
		Table~\ref{tab:ShiftStats}.
		\label{fig:MullikenChargesFigure}
	}
\end{figure}

\begin{figure}
	\begin{center}
		\includegraphics[width=0.5\textwidth]{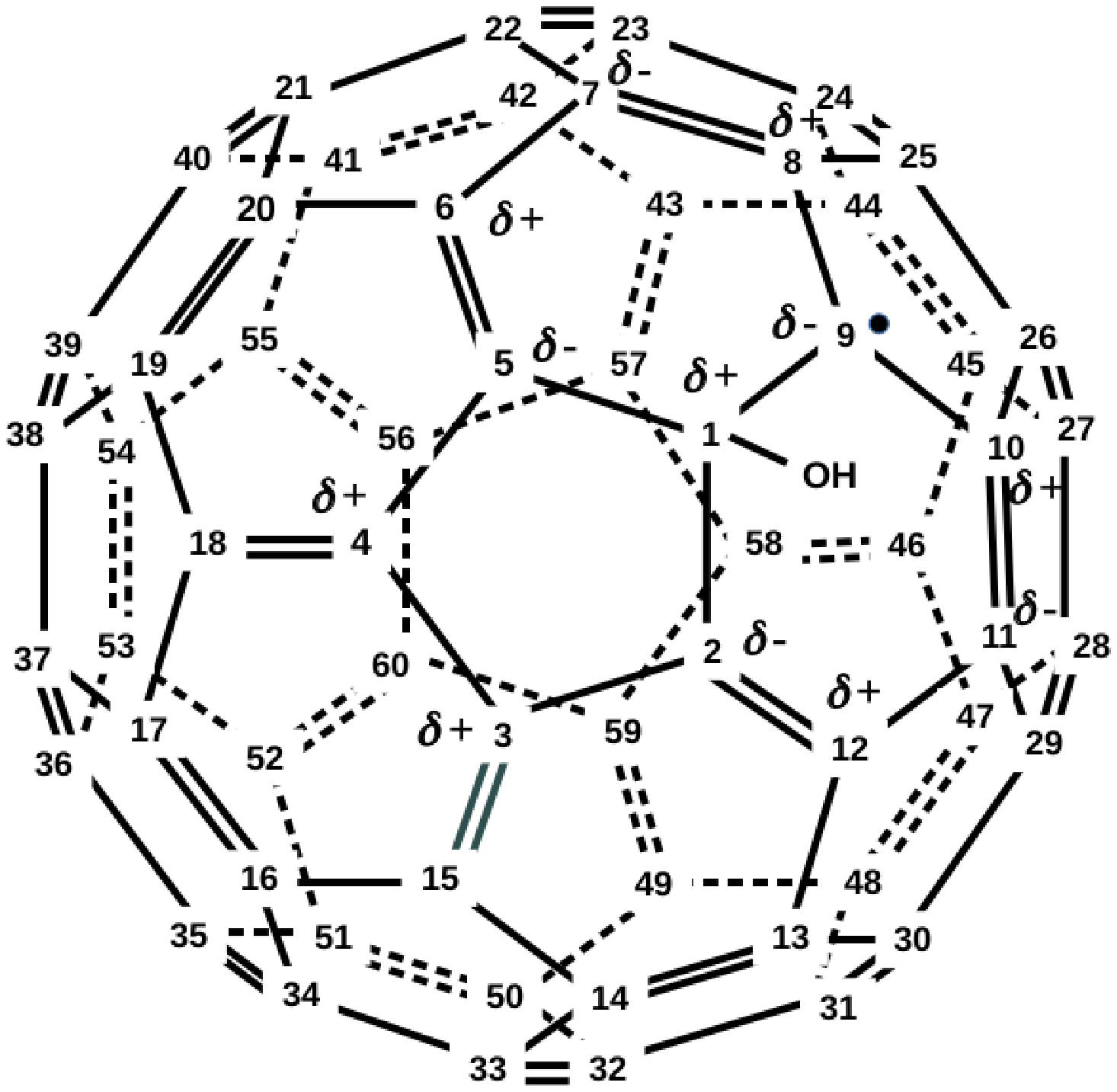}
	\end{center}
	\caption{
		A resonance structure of $^\bullet$C$_{60}$OH with partial charges
		consistent with the Mulliken charges in part (a) of 
		Fig.~\ref{fig:MullikenChargesFigure}.
		\label{fig:ResonanceStructure}
	}
\end{figure}

Since the oxygen is negative in $^\bullet$OH, the Klopman-Salem theory 
also predicts, for the present barrierless reaction, that the BDE should 
grow as the charge of the carbon being attacked grows.  That this is roughly 
true is shown in {\bf Fig.~\ref{fig:MullikenChargesFigure}}.
Looking first at $^\bullet$C$_{60}$OH, we see that the carbon with the
highest BDE is C(9) which is one of the most negatively charged
carbons.  However C(2) and C(5) are also negatively charged and
yet have lower BDEs.  This might be explained by addition at
these sites disrupting 3 hexagons while substitution at C(1) and C(9)
only disrupts 2 hexagons.  Note also that C(2) and C(5) should have
identical BDEs, but only have similar BDEs because different 
conformations have been explored in the two cases.  Interestingly, 
as shown in {\bf Fig.~\ref{fig:ResonanceStructure}}
the most negatively charged carbons are quite close to the most 
positively charged carbons, indicating that charges are formed 
locally with smaller charges on $^\bullet$C$_{60}$OH carbons 
located far from the OH group.  
These charges are relatively easy to rationalize based upon the
idea that oxygen is more electronegative than carbon and the
organic chemistry principles that $sp^2$ carbons are more electronegative
than $sp^3$ carbons, that the radical carbon is $sp^2$ hybridized,
and that the $\pi$ bonds are easily polarized.
For C$_{60}$(OH)$_2$, the most
negatively charged carbons are C(2), C(5), C(8), and C(10).  Up to
some differences in conformation, these are all expected to have
the same charge and the same BDE.  The most positively charged carbons
are C(3), C(4), C(25), and C(26).  By symmetry, up to some conformational
differences, these should all have the same charge and the same
binding energy.  Carbons farther away from the two hydroxyl groups
already present have only small charges.  Note however that in addition to having a more negative charge than C(2), C(8) and C(10), C(5) is the one for which the system has the largest BDE.  Similar observations may
be made for $^\bullet$C$_{60}$(OH)$_3$ and C$_{60}$(OH)$_4$.
In general, many carbons have only small charges, especially when
there are only a few OH groups.  Also there is some scatter due to
not necessarily finding the lowest energy conformer in each case.
{\bf Table~\ref{tab:ShiftStats}} also shows that the BDEs cover a narrower
range when there is an even number of OH groups than when there is
an odd number of OH groups. 
Although there is no theoretical reason to expect a strictly linear 
relationship, we have carried out standard least squares fits of the 
points for each molecule as a convenient way to summarize trends.
We have found these to produce lines with similar slopes.

\begin{table}
	\caption{Average BDE and standard deviation for $^\bullet$OH
		addition to ($^\bullet$)C$_{60}$(OH)$_n$.
		\label{tab:ShiftStats}
	}
	\begin{center}
		\begin{tabular}{ccc}
			\hline \hline
			$n$ & $\langle$ BDE $\rangle \pm \sigma$ (kcal/mol) & Shifted (kcal/mol) `\\
			\hline
			1 & 43.38 $\pm$  7.96 & 0.00 \\
			2 & 47.38 $\pm$  4.89 & 4.00 \\
			3 & 44.47 $\pm$ 11.01 & 1.09 \\
			4 & 44.89 $\pm$  3.93 & 1.51 \\
			5 & 67.59 $\pm$ 10.96 & 24.21 \\
			\hline \hline
		\end{tabular}
	\end{center}
\end{table}

\begin{figure}
	\begin{center}
		\includegraphics[width=0.4\textwidth]{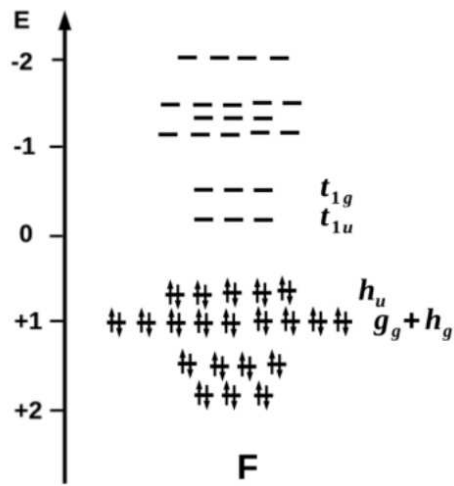}
	\end{center}
	\caption{
		Simple H\"uckel molecular orbital diagram for C$_{60}$ (here designated
		as F for fullerene).
		\label{fig:SHMO}
	}
\end{figure}

On the other hand, the average BDE found for addition to 
$^\bullet$C$_{60}$(OH)$_5$ are considerably larger than for the other
fullerenols (see the statistical data in Table~\ref{tab:ShiftStats}).  
We are not sure why the average BDE should be so much larger for $n=5$, 
but note that a simple H\"uckel molecular orbital theory predicts a five-fold
degenerate HOMO ({\bf Fig.~\ref{fig:SHMO}}).
In order to treat the reactions of the different fullerenols in as equivalent
manner as possible, we have shifted all of the energies to
have the same average BDE as $^\bullet$C$_{60}$(OH) in order to put them
on a single graph (f).  The least squares fit for this graph has a similar
slope and intercept as seen for the graphs of the individual fullerenols.
On the other hand, the points are highly scattered with a very low 
coefficient of determination ($R^2$) near only 0.3.  This is partly
explained by not always finding the minimum energy conformation for
each isomer and by the low value of atomic charges for most of the carbons.
Nevertheless, the trends are clear.
We have no reason to expect the linearity of the points to be improved by
using some other definition of atomic charges.  This is probably the best that
can be done when it comes to predicting the reactivity of C$_{60}$(OH)$_n$ when
$n$ is even.

\begin{figure}
	\begin{center}
		\includegraphics[width=1.2\textwidth]{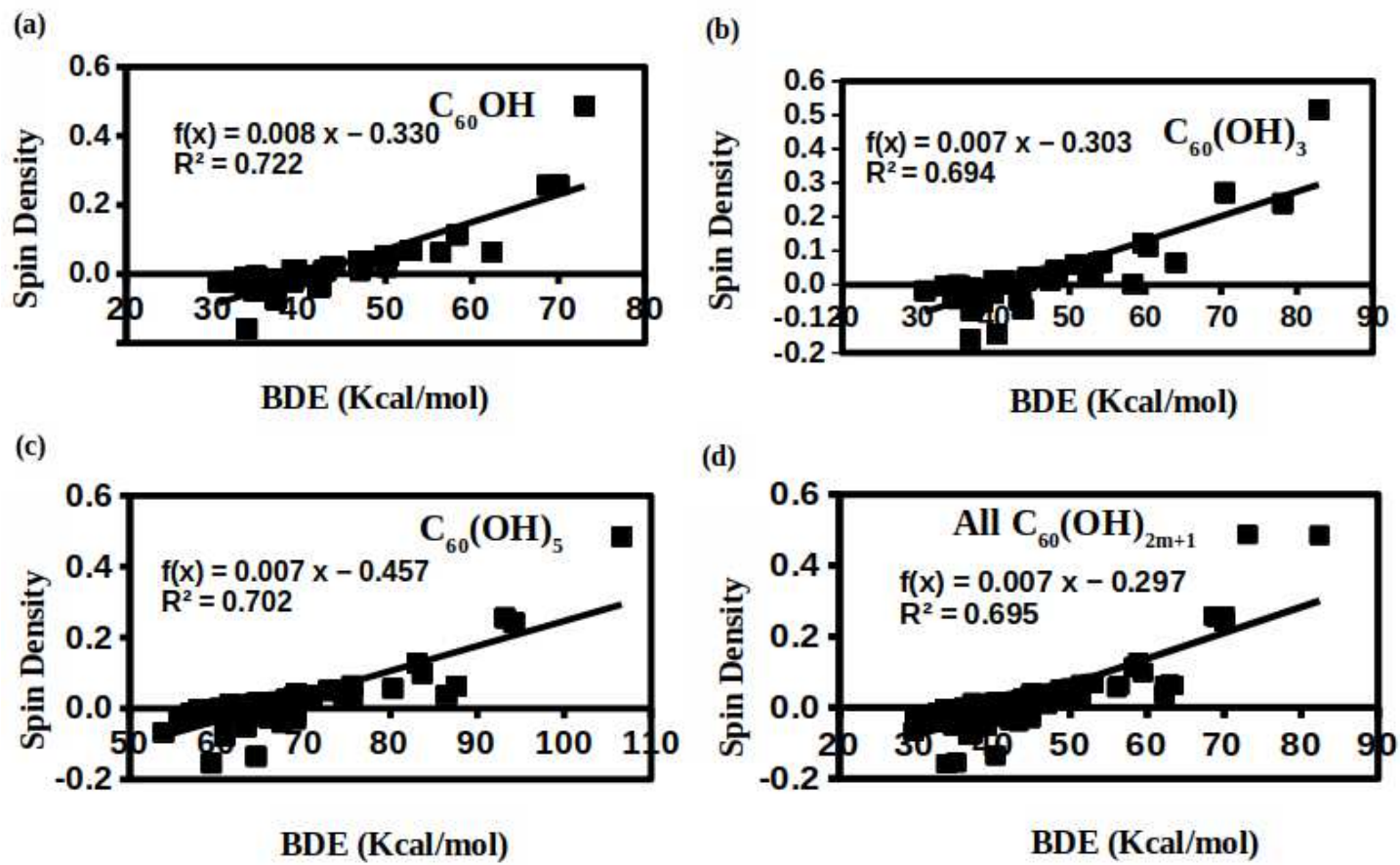}
	\end{center}
	\caption{
		Correlation between carbon spin-densities and associated BDEs.
		The BDE in part (d) is shifted using the statistical data in 
		Table~\ref{tab:ShiftStats}.
		\label{fig:SpinDensitiesFigure}
	}
\end{figure}

A common belief, no doubt based upon reasonable physical intuition, is
that the regioselectivity of radical species in radical-radical reactions should
take place where the spin-density $\rho_\uparrow-\rho_\downarrow$ is greatest.
{\bf Figure~\ref{fig:SpinDensitiesFigure}} shows that this idea is essentially
correct, though it cannot be extended to molecules with an even number of
hydroxyl groups as our closed-shell calculations are spin-restricted with
spin-densities which are rigorously zero. This would seem to imply a certain
amount of orbital-control. 
The correlation (as measured by a least-squares-fit $R^2$ value of about 0.7)
is significantly better than for the Mulliken charges.  Once again, we may
place all the data on a single graph (d) with similar slope and intercept.

\begin{figure}
	\begin{center}
		\includegraphics[width=1.2\textwidth]{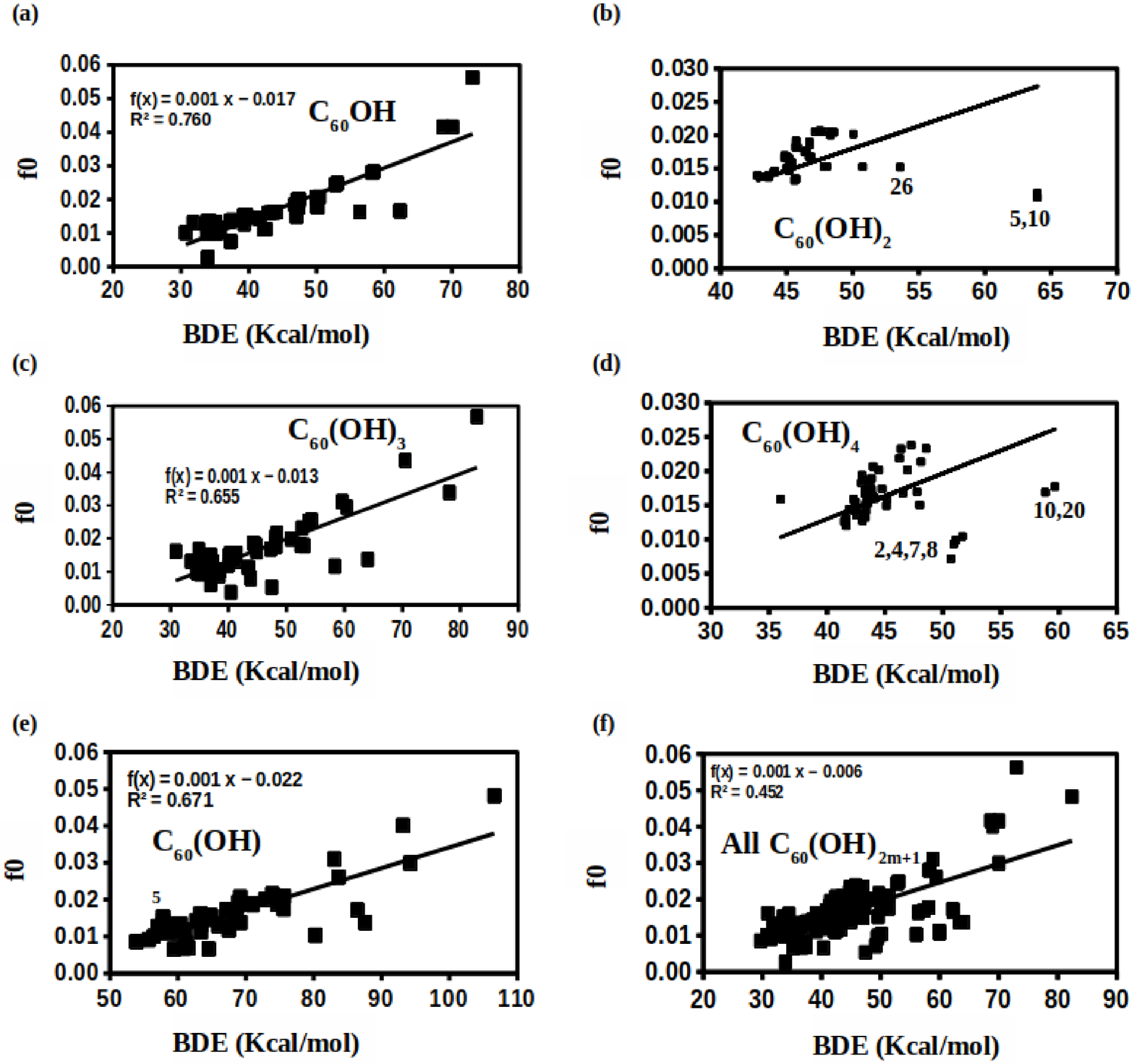}
	\end{center}
	\caption{
		Correlation between carbon CFFs and associated BDEs.  Some points have been labeled by their carbon number to facilitate discussion.  The BDE in part (f) is shifted using the statistical data from
		Table~\ref{tab:ShiftStats}.  The lines in parts (b) and (d) are
		$y=m(x-\Delta x)+b$ where $y=mx+b$ is the least-squares-fit line
		obtained in part (f) and $\Delta x$ is the shift from
		Table~\ref{tab:ShiftStats}.
		\label{fig:f0CFF}
	}
\end{figure}

CFFs offer some advantages over the spin-density.  For one thing, the
CFFs may be calculated for both open- and closed-shell molecules.
Radical reactivity is associated with the $f_0$ CFF.  This quantity
is shown in {\bf Figure~\ref{fig:f0CFF}}.  
For fullerenols with an odd number of hydroxyl groups, 
Fig.~\ref{fig:f0CFF} shows that there is an excellent correlation between 
the $f_0$ CFF and the BDE.  The quality of the correlation is
similar ($R^2 \approx 0.7$) as for the spin-density.
For fullerenols with an even number of hydroxyl groups, the graph is
much more scattered.  Nevertheless many points are actually quite close
to the least-square fit line obtained from the fullerenols with an
odd number of hydroxyl groups.  This observation could not have been
made for the spin-density as the spin-density is zero in the case of
fullerenols with an even number of hydroxyl groups.

\begin{figure}
	\begin{center}
		\includegraphics[width=1.2\textwidth]{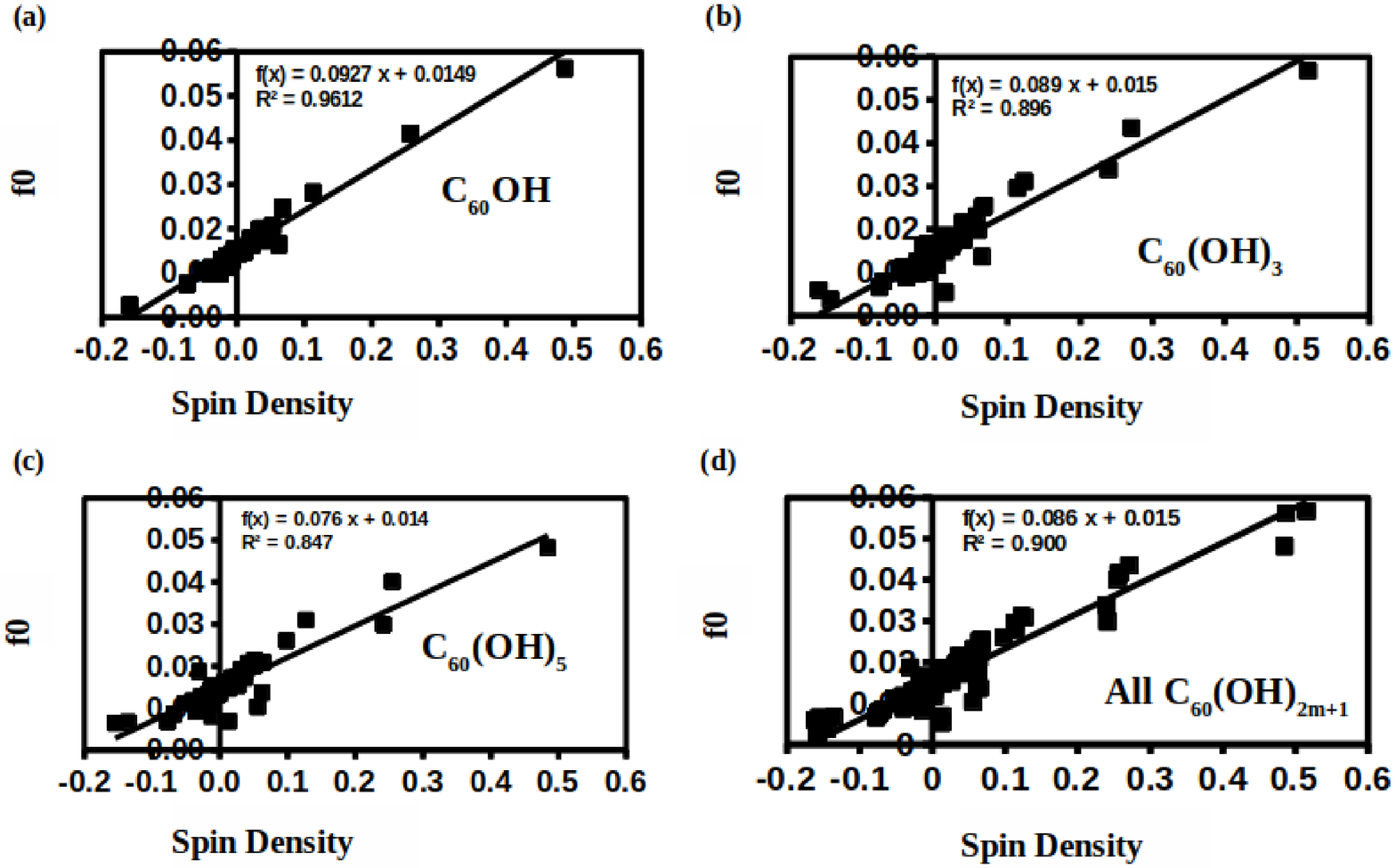}
	\end{center}
	\caption{
		Correlation between the $f^0$ CFF and the spin-density for radical fullerenols.
		\label{fig:f0SpinDensity}
	}
\end{figure}

A question that we may then ask is why the spin-density and $f_0$ behave
so similarly?  In the case of the radical fullerenols, the three Fukui 
functions are all approximately the SOMO density $\vert \psi_S \vert^2$ 
which is also the only place available to put the excess spin in the frozen
orbital approximation.  Hence we expect $f^0$ and the spin-density to be
nearly identical.  {\bf Figure~\ref{fig:f0SpinDensity}} shows that this is
roughly the case.  Of course, they are not exactly identical because the
frozen orbital approximation is not strictly valid.

\begin{figure}
	\begin{center}
		\includegraphics[width=1.2\textwidth]{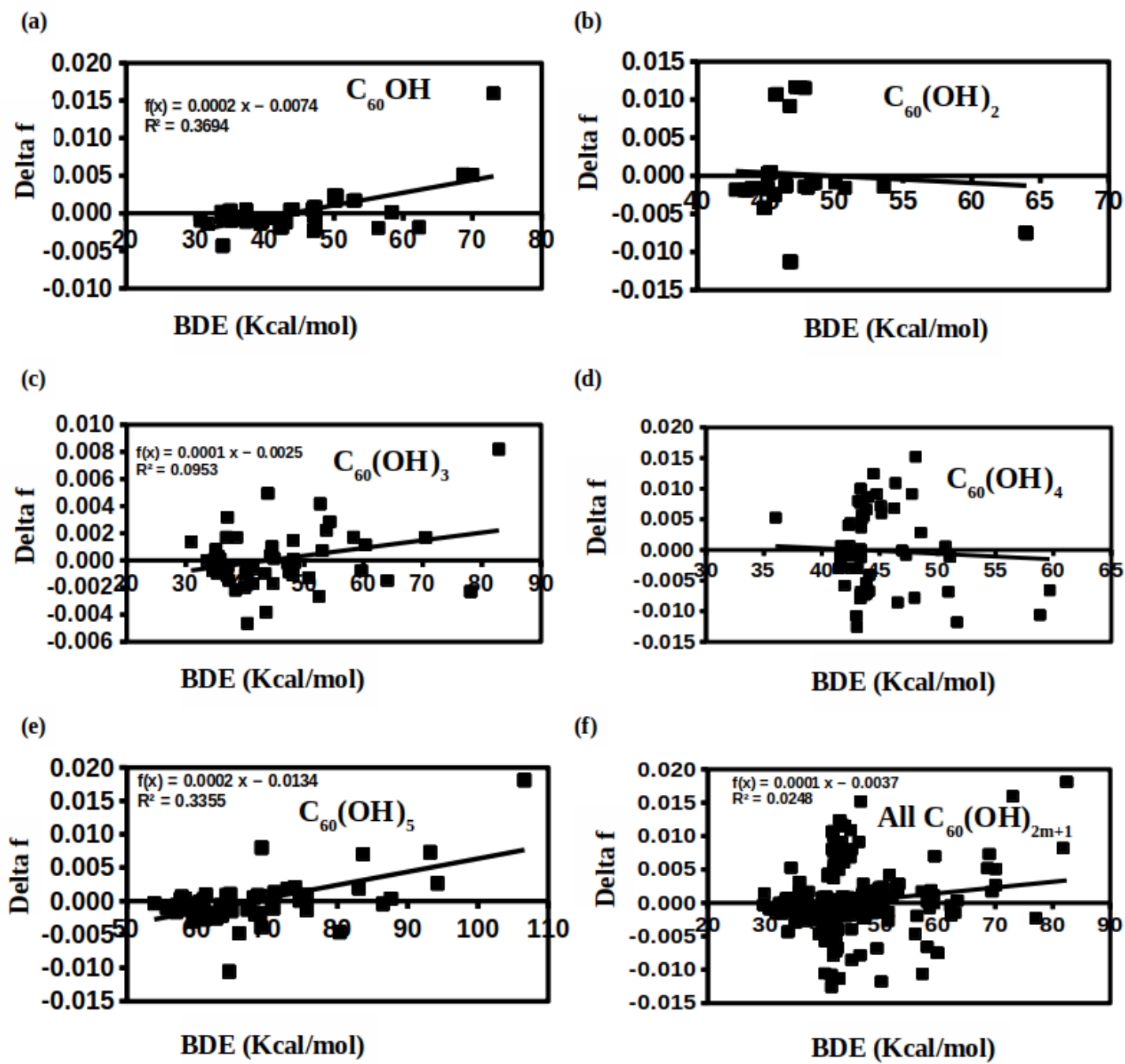}
	\end{center}
	\caption{
		Correlation between carbon $\Delta f$ and associated BDEs.
		The BDE in part (f) is shifted using the statistical data from
		Table~\ref{tab:ShiftStats}.  The lines in parts (b) and (d) are
		$y=-[m(x-\Delta x)+b]$ where $y=mx+b$ is the least-squares-fit line
		obtained in part (f) and $\Delta x$ is the shift from 
		Table~\ref{tab:ShiftStats}.
		\label{fig:DualDescriptorFigure}
	}
\end{figure}

\begin{figure}
	\begin{center}
		\includegraphics[width=0.8\textwidth]{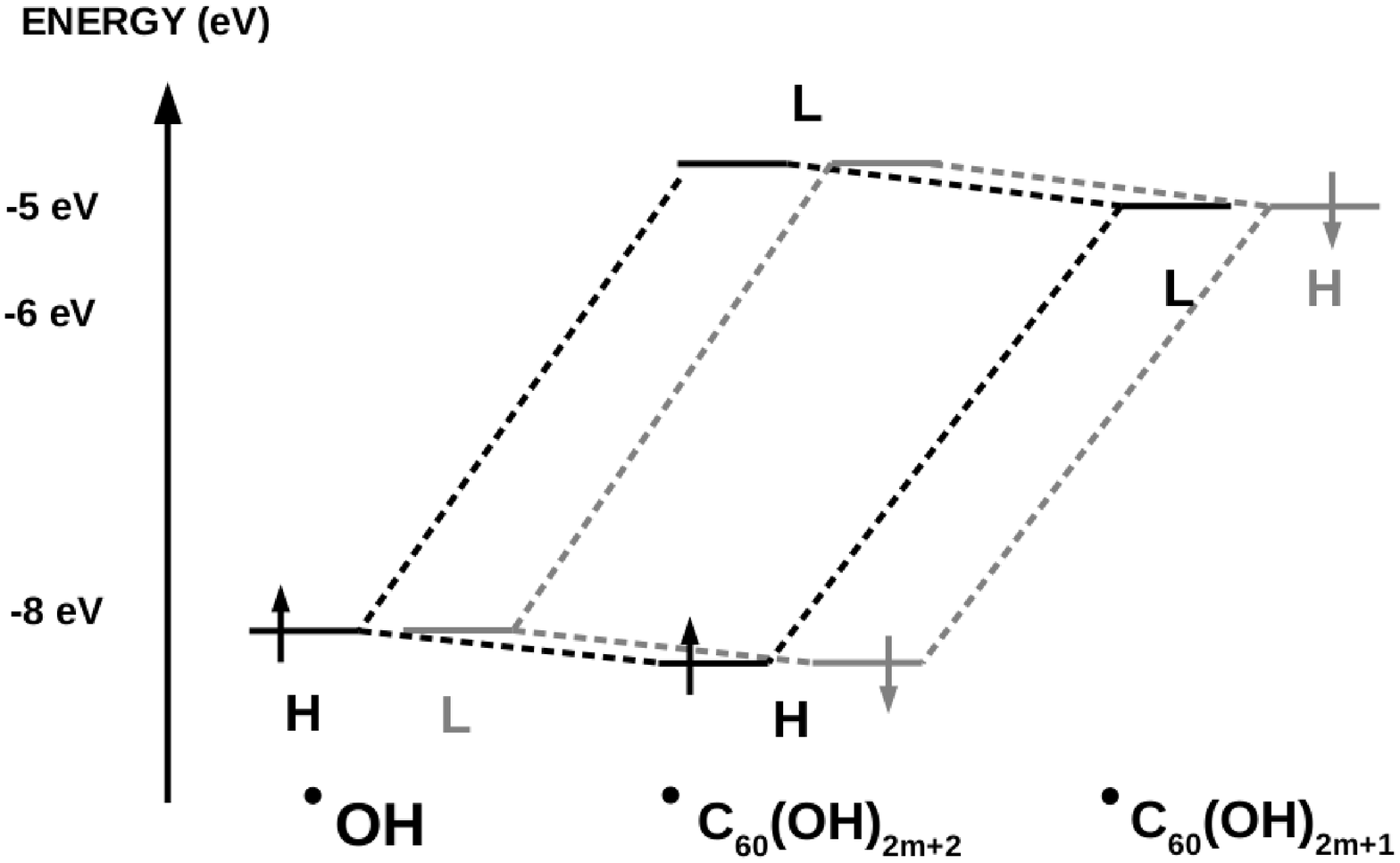}
	\end{center}
	\caption{
		Radical-radical FMO correlation diagram.
		\label{fig:RadialRadicalFMO}
	}
\end{figure}

We may now examine the claim that $^\bullet$OH is strongly electrophillic
\cite{MIN+02} to see what, if any, validity it has for regioselectivity.  
{\bf Figure~\ref{fig:DualDescriptorFigure}} shows how the dual descriptor 
$\Delta f$ varies with BDE. The expectation, based upon the usual
FMO diagram (Fig.~\ref{fig:FMOcorrelationdiag}) is that the largest BDE
should be where the HOMO is largest and hence where the dual descriptor
is most negative.  In fact, the figure shows the opposite trend for the
reaction of $^\bullet$OH with radical fullerenols.  The resolution of this
conundrum is shown in {\bf Fig.~\ref{fig:RadialRadicalFMO}} where we see
that the spin up HOMO of $^\bullet$OH is actually interacting with the
spin down LUMO of $^\bullet$C$_{60}$(OH)$_{2m+1}$.  Thus the largest
BDE is actually where the LUMO dominates which corresponds to a negative
dual descriptor.  Figure~\ref{fig:DualDescriptorFigure} also shows that
the negative of the least squares fit line for the odd fullerenols also
gives the rough slope of the data for the even fullerenols, confirming
that $^\bullet$OH is also acting like an electrophilic radical in these
cases.

\section{Concluding Discussion}

The hydroxyl radical $^\bullet$OH is a highly reactive oxygen species 
and buckminsterfullerene C$_{60}$ is a radical sponge.   They are known 
to react easily with each other.  This has pros and cons.  On the negative
side, most radical reactions are relatively unselective and therefore of
little use to synthetic chemists.  On the positive side, C$_{60}$ may 
be an excellent trap for removing $^\bullet$OH.

It is known that ($^\bullet$)C$_{60}$(OH)$_n$ isomers differ in their 
relative stabilities, with the most stable isomers formed by
successive addition of C$_{60}$ by $^\bullet$OH having 
a ring of OH groups around the equator of C$_{60}$.  The present work
has been aimed at answering the question, ``Why are some isomers
more stable than others?'', using the tools of frontier molecular 
orbital theory.  In particular, we used the concepts
of Fukui functions and dual descriptors from conceptual density-functional
theory as well as spin-densities to try to understand regioselectivity.  
Localized (i.e., condensed atomic) quantities were calculated using 
Mulliken population analysis.  Results were analyzed in terms of the 
Klopman-Salem interpretation of frontier molecular orbital theory.

We call this a laboratory for testing frontier molecular orbital theory
for several reasons.  The first is because the high symmetry, high-density
of states, and frequent open-shell nature of ($^\bullet$)C$_{60}$(OH)$_n$ 
means that successive addition of C$_{60}$ by $^\bullet$OH is a prime
candidate for the breakdown of some of the assumptions behind the 
Klopman-Salem interpretation of frontier molecular orbital theory.  
A second reason is that each ($^\bullet$)C$_{60}$(OH)$_n$ structural
isomer also contains a number of different conformers.  No attempt was
made to search for the absolute minimum energy conformer in each case
which leads to errors as large as 5 kcal/mol in bond dissociation energies
for each successive addition.
Thirdly, the choice of Mulliken population analysis may be questioned, though
our basis is well balanced and the Mulliken approach has the advantage of
being the oldest, and hence best understood, localization method.  The result
of these three caveats is that there is a great deal of scatter in our results.
This is countered by the fact that we have carried out calculations for
285 reactions, which is enough for clear trends to emerge.

Plots against bond dissociation energies are not generally linear because
reaction energies far from pre-existant hydroxyl sites are expected to
be relatively constant while much greater variation in reaction energies
is expected close to other hydroxyl groups.  Nevertheless, for lack of
a better tool, we have performed least square fits of these graphs as
an aid to the quantitative discussion of trends.  Other graphs such
as that of the condensed radical Fukui function $f^0$ against spin density
are expected to be roughly linear and this is exactly what has been found.

Klopman-Salem frontier molecular orbital theory applies best to the analysis
of bond dissociation energies for reactions with an early transition state.
This is the case for the gas-phase reactions studied here in the sense that
no transition barriers have been encountered.  In the Klopman-Salem theory,
the reaction energy is further decomposed into a steric term, a charge-control
term, and an orbital-control term.  We neglect the steric term in our study
because the main steric contribution would come from hydrogen bonds which
are already considered in the 5 kcal/mol differences in conformer energies.
The separation between charge-control and orbital-control turns out to be
far less obvious.  The energy gap between the frontier molecular orbitals
of $^\bullet$OH and the fullerenols is large enough to expect significant
charge control.  However, especially in the case of open-shell fullerenols,
substantial orbital control is also expected to occur.  These observations
are nicely summarized by a rough frontier molecular orbital diagram which 
predicts that $^\bullet$OH should be an electrophilic radical.

Odd $^\bullet$C$_{60}$(OH)$_{2m+1}$ and even C$_{60}$(OH)$_{2m}$ fullerenols
are found to have very different reactivities towards $^\bullet$OH.  In 
particular, there is much less spread in calculated bond dissociation energies
for even fullerenols than for odd fullerenols, showing the enormous importance
of the unpaired electron in $^\bullet$C$_{60}$(OH)$_{2m+1}$.  This shows up
in the strong, relatively linear, correlation of the spin density with 
bond dissociation energies in this case.  However the condensed radical
Fukui function $f^0$ correlates strongly with the spin density and may be
used, not just to explain trends in the bond dissociation energies of 
the odd $^\bullet$C$_{60}$(OH)$_{2m+1}$ fullerenols, but also also to provide 
general trends for the even C$_{60}$(OH)$_{2m}$ fullerenols for which the
spin-density is zero.  This shows a clear advantage of conceptual 
density-functional theory ideas over simply looking at charges and
spin densities and is a form of orbital control.  However plots of 
Mulliken charge densities against bond dissociation energies also show
a strong preference for binding at sites with negative carbons, independent
of whether the fullerenol has an even or an odd number of hydroxl groups,
which confirms the importance of charge control.

The electrophilic nature of $^\bullet$OH was further confirmed by calculations
of the dual descriptor $\Delta f$.  Usually the electrophilic nature of 
$^\bullet$OH would be expected to be associated with a negative slope of
a graph of the dual descriptor versus the bond dissociation energy.  This
trend is found for even C$_{60}$(OH)$_{2m}$ fullerenols.  Interestingly
a negative slope is found for the same type of graph for odd 
$^\bullet$C$_{60}$(OH)$_{2m+1}$ fullerenols.  The reason for this 
unexpected result lies in the nature of the open-shell spin-unrestricted
density-functional theory calculations themselves where the definitions of the 
relevent HOMO and LUMO spin-orbitals differ from the expected definitions
used in spin-restricted density-functional theory.

Frontier molecular orbital theory provides a clear and useful picture of
the reactivity of hydroxyl radicals with buckminsterfullerene.  However it
should be kept in mind that it is not the only analytical tool.  Further
information is likely to be obtained by examining aromaticity criteria,
notably by counting Clar structures.  This is the most likely explanation
of why the hydroxyl radicals substitute first around the equator of C$_{60}$.
Further information about why the Mulliken charges are the way they are
can be obtained from induction and electronegativity arguments.  
Although not emphasized, these ideas have been alluded to in the 
text.  As our ultimate objective is to understand how buckminsterfullerene can
act as a radical scavanger on the skin and within cells, we are continuing
our work using using solvation models.

\section{Acknowledgements}

The authors wish to acknowledge the support from {\em l'Universit\'e Grenoble Alpes}'s
ICMG ({\em Institut de Chimie Mol\'eculaire de Grenoble}) Chemistry Nanobio Platform
PCECIC ({\em Plateau du Centre d'Exp\'erimentation et de Calcul Intensif en Chimie})
on which this work has been performed.  Pierre GIRARD is gratefully acknowledged
for his help and support using this platform. Andr\'es Cisneros, Paul Geerlings, Henry Chermette,  Christophe Morell and Goedele Roos  are acknowledged for useful discussions.

\section*{Supplementary Information}
\begin{itemize}
	\item CRediT contributor roles \cite{CRediT}.
	\item Basis Set Validation
	\item Tables of Bond Dissociation Energies, Spin Densities, Mulliken, and Condensed Fukui Functions
\end{itemize}

\end{document}